\documentclass[12pt]{article}

\usepackage{graphicx}
\usepackage{caption}
\usepackage{times}
\usepackage{wasysym} 
\usepackage[superscript]{cite}
\usepackage[T1]{fontenc}

\newcommand{\onlinecite}[1]{\hspace{-1 ex} \nocite{#1}\citenum{#1}} 

\newcommand{\yto}{{Yb$_2$Ti$_2$O$_7$}\ }

\title{{\Huge Supplementary Information}}
\date{}

\begin{document}

\baselineskip30pt

\noindent{\Large{\textbf{{\textsf{Quantum Excitations in Quantum Spin Ice}}}}}
\bigskip

\noindent Kate A. Ross$^{1}$, Lucile Savary$^2$, Bruce D. Gaulin$^{1,3,4}$ \& Leon Balents$^{5,*}$

\bigskip

\bigskip

{\sl\noindent$^1$Department of Physics and Astronomy, McMaster University, Hamilton, Ontario, L8S 4M1, Canada\\
$^2$Ecole Normale Sup\'{e}rieure de Lyon, 46, all\'{e}e d'Italie, 69364 Lyon Cedex 07, France\\
$^3$Canadian Institute for Advanced Research, 180 Dundas St.\ W., Toronto, Ontario, M5G 1Z8, Canada\\
$^4$Brockhouse Institute for Materials Research, McMaster University, Hamilton, Ontario, L8S 4M1, Canada\\
$^5$Kavli Institute for Theoretical Physics, University of California, Santa Barbara, CA, 93106-4030, U.S.A.\\}

\noindent$^*$ corresponding author, email: balents@kitp.ucsb.edu

\oddsidemargin=-.0cm
\evensidemargin=-.0cm
\textwidth=16.5cm
\topmargin=-1.35cm
\textheight=22cm

\newpage

\noindent{\bf 
  Recent work has highlighted remarkable effects of classical thermal
  fluctuations in the dipolar spin ice compounds, such as ``artificial
  magnetostatics'', manifesting as Coulombic power-law spin correlations
  and particles behaving as diffusive ``magnetic monopoles''.  In this
  paper, we address {\sl quantum} spin ice, giving a unifying framework
  for the study of magnetism of a large class of magnetic compounds with
  the pyrochlore structure, and in particular discuss Yb$_2$Ti$_2$O$_7$
  and extract its full set of Hamiltonian parameters from high field
  inelastic neutron scattering experiments.  We show that fluctuations
  in Yb$_2$Ti$_2$O$_7$ are strong, and that the Hamiltonian may support
  a Coulombic ``Quantum Spin Liquid'' ground state in low field and host
  an unusual quantum critical point at larger fields.  This appears
  consistent with puzzling features in prior experiments on Yb$_2$Ti$_2$O$_7$.  Thus
  \yto is the first quantum spin liquid candidate in which
  the Hamiltonian is quantitatively known.
}

\bigskip

Rare earth pyrochlores display a diverse set of fascinating physical
phenomena.\cite{gardner2010magnetic} One of the most interesting
aspects of these materials from the point of view of fundamental
physics is the strong frustration experienced by coupled magnetic
moments on this lattice.  The best explored materials exhibiting this
frustration are the ``spin-ice'' compounds, Ho$_2$Ti$_2$O$_7$,
Dy$_2$Ti$_2$O$_7$, in which the moments can be regarded as {\sl
  classical} spins with a strong easy-axis (Ising)
anisotropy.\cite{bramwell2001spin, clancy2009revisiting} The
frustration of these moments results in a remarkable {\sl classical
  spin liquid} regime displaying Coulombic correlations and emergent
``magnetic monopole'' excitations that have now been studied
extensively in theory and experiment.\cite{castelnovo2008magnetic,
  fennell2009magnetic, jaubert2009signature}

Strong quantum effects are absent in the spin ice compounds, but can
be significant in other rare earth pyrochlores.   In particular, in many
materials the low energy spin dynamics may be reduced to that of an
effective spin $S=1/2$ moment, with the strongest possible quantum
effects expected.  In this case symmetry considerations reduce the exchange constant
phase space of the nearest neighbour exchange Hamiltonian to a maximum
of three dimensionless parameters.\cite{PhysRevB.78.094418} The
compounds Yb$_2$Ti$_2$O$_7$, Er$_2$Ti$_2$O$_7$, Pr$_2$Sn$_2$O$_7$\cite{gardner2010magnetic} (and
possibly 
Tb$_2$Ti$_2$O$_7$\cite{molavian-gingras-canals}) are of this
type, and it has recently been
argued that the spins in \yto and Er$_2$Ti$_2$O$_7$ are controlled by
exchange coupling rather than by the long-range dipolar interactions
which dominate in spin ice.\cite{citeulike:8803175,thompson2010rods}
This makes these materials beautiful examples of strongly quantum
magnets on the highly frustrated pyrochlore lattice.  They are also
nearly ideal subjects for detailed experimental investigation,
existing as they do in large high purity single crystals, and with
large magnetic moments amenable to neutron scattering studies.  \yto
is particularly appealing because its lowest Kramers doublet is
extremely well separated from the first excited one,\cite{Hodges2001}
and a very large single crystal neutron scattering data set is
available, allowing us to measure the full Hamiltonian quantitatively,
as we will show.  Although we specialize to the latter material in the
present article the theoretical considerations and parameter
determination method described here will very generally apply to {\sl
  all} pyrochlore materials where exchange interactions dominate and
whose dynamics can be described by that of a single doublet.


Theoretical studies have pointed to the likelihood of unusual ground
states of quantum antiferromagnets on the pyrochlore lattice.\cite{canals1998pyrochlore,PhysRevB.69.064404}  Most
exciting is the possibility of a {\sl quantum spin liquid} (QSL) state,
which avoids magnetic ordering and freezing even at absolute zero
temperature, and whose elementary excitations carry fractional quantum
numbers and are decidedly different from spin waves.\cite{balents2010spin}  Intriguingly, neutron scattering measurements
have reported a lack of magnetic ordering and the absence of spin waves
in \yto at low fields.\cite{hodgesfluc,ross2009}  In a recent study,
sharp spin waves emerged when a magnetic field of $0.5$T or larger was
applied, suggesting that the system transitioned into a conventional
state.\cite{ross2009}  The possible identification of the low field
state with a quantum spin liquid is tantalizing, but progress certainly
requires a more detailed understanding of the spin Hamiltonian.

\bigskip

In this article, we present a detailed experimental and theoretical
investigation of the excitation spectrum in the high field state
throughout the Brillouin zone.  We show that the spectrum is extremely
well fit by spin wave theory, and through this fit we unambiguously
extract {\sl all} the microscopic exchange parameters in
the spin Hamiltonian (see below).  Interestingly, we find that the
largest exchange interaction is of the same ``frustrated ferromagnetic''
Ising type as in spin ice, despite the fact that the g-tensor tends to
orient magnetic moments primarily normal to the Ising axes.  Moreover,
spin-flip terms which induce quantum dynamics are comparable to the
Ising exchange, which confirms the picture of \yto as a strongly quantum
magnet, in qualitative agreement with recent studies.\cite{cao2009aniso,
  thompson2010rods,Onoda} Strikingly, we find that the predictions of
mean field theory (MFT) using these parameters disagree drastically with
experiment in zero field, indicating that fluctuations strongly
reduce or destroy the classically expected spin order.  Taken together,
these observations make \yto a promising candidate for observation of
QSL physics.  The precise determination of the microscopic spin
interactions sets the stage for a quantitative understanding and test of
this proposal.

\bigskip

Time-of-flight neutron scattering measurements were performed on a 7g
single crystal of Yb$_2$Ti$_2$O$_7$, grown via the optical floating zone
method.  Details of the crystal growth were given elsewhere.\cite{ross2009,gardner_growth}  The neutron scattering data was
collected at the Disk Chopper Spectrometer at the NIST Center for
Neutron Research, using $5$\AA \ incident neutrons.  This configuration
allowed an energy resolution of $0.09$meV. The sample environment
consisted of a 11.5T vertical field magnet combined with a dilution
insert that achieved a base temperature of 30mK.  The scattering plane
was HHL, with the field applied along the [1$\bar{1}$0] vertical
direction.  The sample was rotated $147$ degrees in $1.5$ degree steps about
the vertical, allowing a three dimensional data set to be acquired,
i.e. two components of the wavevector \textbf{Q} within the scattering
plane, and energy transfer.  The spin excitation spectra along several
high symmetry directions with the scattering plane were thereby obtained.  

At $30$mK, the inelastic spectrum changes qualitatively at $H=0.5$T; above this field strength, resolution-limited spin wave excitations which go soft with quadratic dispersion at nuclear-allowed positions develop, indicating a transition to a field-polarized ferromagnetic state.\cite{ross2009}  The spin wave excitations indicate that the symmetry of the underlying lattice is preserved, as is evident from gaps in the spectrum at the nuclear zone boundaries.  In Figure 1 we show the spin wave dispersions along several directions in the HHL plane for $H=2$T and  $H=5$T.   These high symmetry directions are shown relative to the Brillouin zone structure within the HHL plane in Fig. 2.





\bigskip

We compare the experimental data to spin-wave theory.  We assume
nearest-neighbour exchange coupling only, as appropriate to the strongly
localized $f$-electron states, and neglect dipolar interactions.
The Hamiltonian, written in global spin coordinates, is then
\begin{equation}
  \label{eq:1}
  H = \frac{1}{2}\sum_{ij} J_{ij}^{\mu \nu} S_i^\mu S_j^\nu - \mu_B H^\mu \sum_i
  g_i^{\mu \nu} S_i^\nu,
\end{equation}
where $J_{ij}^{\mu\nu} = J_{ji}^{\nu\mu}$ is the matrix of exchange
couplings between sites $i$ and $j$, $g_i^{\mu\nu}$ is the g-tensor
for the spin at site $i$, and we take $\hbar=1$.  Symmetry allows four independent exchange
constants,\cite{PhysRevB.78.094418} $J_1,\cdots,J_4$.  To specify them, we give the
exchange matrix on one pair of nearest-neighbour sites, located at
positions ${\bf r}_0 = \frac{a}{8}(1,1,1)$ and ${\bf r}_1=\frac{a}{8} (1,-1,-1)$
on a tetrahedron centered at the origin ($a$ is the conventional cubic
lattice spacing for the fcc Bravais lattice):
\begin{equation}
  \label{eq:2}
  \mathbf{J}_{01} = \left( \begin{array}{ccc} J_2 & J_4 & J_4 \\ -J_4 & J_1 &
      J_3 \\ -J_4 & J_3 & J_1 \end{array} \right).
\end{equation}
The other exchange matrices can be obtained from this one by cubic
rotations given in the Supplementary Information.  The g-tensor contains two components: $g_z$ parallel to and
$g_{xy}$ perpendicular to the {\sl local} $C_3$ rotation axis through
the Yb site.  


\bigskip

Spin wave theory, carried out as described in the Supplementary Information, is fit to the $H = 5\,$T, $T = 30\,$mK
measurements; the fitting procedure focuses on the dispersion relation
alone, and the overall intensity of the calculated spin waves is scaled
to agree with the experiment at a single wavevector and energy
point. The resulting inelastic structure factor $S(\mathbf{Q},\omega)$ (see Supplementary Information) is compared to both the $5$T and $2$T data in Figure
1.  The best fit is achieved with the following exchange
parameters, in meV:
\begin{equation}
  \label{eq:fitparams}
  J_{1} = -0.09 \pm 0.03,\; J_{2} = -0.22 \pm 0.03,\; J_{3}=-0.29\pm
  0.02 ,\; J_{4} = 0.01 \pm 0.02.
\end{equation}
Here we quote rough uncertainties obtained by the visual
comparison of the theoretical and experimental intensities.  The fit is performed by taking
the ratio of components of the g-tensor to be $g_{xy}/g_z = 2.4$,
i.e. the ratio obtained by Ref. \onlinecite{Hodges2001}. The fit then produces $g_z=1.80$, in nearly
perfect agreement with these studies (using the g-factor ratio of Cao
{\sl et al.} instead,\cite{cao2009aniso} i.e. $g_{xy}/g_z$ = 1.8, does not reproduce the data as precisely).  Using these results, a
high temperature expansion gives (see Supplementary Information) a theoretical Curie-Weiss temperature
$\Theta_{CW}=312$mK, which is comparable to but smaller than the
experimentally determined values, $\Theta_{CW}=400$mK\cite{Blote1969}
and $750$mK\cite{Hodges2001}.  The deviations may be explained by the
sensitivity of the theoretical value to small changes in the g-factors
and exchange parameters, and to the dependence of the experimental value
on the fitting range.\cite{cao2009aniso}  Furthermore, and most
importantly, our extracted exchange parameters correctly reproduce
relative intensities as well as the shape of the spin wave dispersion
for each of the five directions.  Agreement is excellent for $H=2$T, showing that these parameters produce a robust description of
the field-induced ferromagnetic state.  We note, however, that there is a
significant quantitative disagreement with the exchange parameters
quoted in Refs. \onlinecite{citeulike:8803175,thompson2010rods} (see
Supplementary Information).

{\sl Implications: } The excellent agreement with spin-wave theory for
fields $H \geq 2$T clearly indicates that the high field state is
accurately modeled semi-classically, and is smoothly connected to the
fully polarized limit.   Theoretically, the ground state in this regime
breaks no symmetries, and supports a ferromagnetic polarization along
the axis of the applied field (for the $\langle 110\rangle$ field used
in the experiment).  However, the semiclassical analysis clearly and
dramatically fails at small fields, where the measurements show no signs of spontaneous long range order.\cite{ross2009}  The classical zero field ground
state for our Hamiltonian parameters has a large spontaneous
polarization along the $\langle 100\rangle$ axis.  Extension of this
analysis to a $T>0$ mean-field theory wrongly predicts a continuous
magnetic ordering transition at a temperature of $T_c^{MF} =
3.2$K (see Supplementary Information). The experimental indications of a zero-field transition to long range order are mixed,\cite{yasui,gardnerpol} but early
specific heat measurements\cite{Blote1969} found an anomaly at
$T_{c}=214$mK, and M\"{o}ssbauer spectroscopy\cite{Hodges2002}
suggested a transition at $240$mK.  This temperature is approximately
14 times lower than $T_c^{MF}$.  If there is magnetic ordering at all,
it appears to be substantially suppressed, indicating strong
fluctuations -- classical, quantum, or both.

\bigskip

The presence of strong fluctuations makes a QSL ground state plausible 
in low field.  We now use the Hamiltonian parameters to suggest the
nature of this state.  To do so, we rewrite the zero field Hamiltonian
in terms of spins quantized along the local $C_3$ axis for each site,
similarly to Ref. \onlinecite{Onoda}:
\begin{eqnarray}
  \label{eq:4}
  H & = & \sum_{\langle ij\rangle} \Big\{ J_{zz} \mathsf{S}_i^z \mathsf{S}_j^z - J_{\pm}
  (\mathsf{S}_i^+ \mathsf{S}_j^- + \mathsf{S}_i^- \mathsf{S}_j^+) \nonumber  + J_{\pm\pm} \left[\gamma_{ij} \mathsf{S}_i^+ \mathsf{S}_j^+ + \gamma_{ij}^*
    \mathsf{S}_i^-\mathsf{S}_j^-\right] \nonumber \\
&& + J_{z\pm}\left[ \mathsf{S}_i^z (\zeta_{ij} \mathsf{S}_j^+ + \zeta^*_{ij} \mathsf{S}_j^-) +
  {i\leftrightarrow j}\right]\Big\},
\end{eqnarray}
where here $\mathsf{S}_i^\mu$ are {\sl local} spin coordinates, $J_{zz} = -\frac{1}{3}(2J_1-J_2+2(J_3+2J_4))$, $J_{\pm}=\frac{1}{6}(2J_1-J_2-J_3-2J_4)$, $J_{\pm\pm}=\frac{1}{6}(J_1+J_2-2J_3+2J_4)$ and $J_{z\pm}=\frac{1}{3\sqrt{2}}(J_1+J_2+J_3-J_4)$, and the matrices $\gamma_{ij}, \zeta_{ij}$ consist
of unimodular complex numbers (given in the Supplementary Information).
From the fits in Eq.~(\ref{eq:fitparams}), we find, in meV,
\begin{equation}
  \label{eq:5}
  J_{zz} = 0.17\pm 0.04,\; J_{\pm} = 0.05\pm 0.01,\;
  J_{\pm\pm}=0.05\pm 0.01,\; J_{z\pm} = - 0.14\pm 0.01,
\end{equation}
where the uncertainties have been estimated by treating those in
Eq.~(\ref{eq:fitparams}) as Gaussian random variables.
Note that the strongest interaction is $J_{zz}>0$, which precisely
coincides with the nearest-neighbour spin-ice model.  The model with
$J_\pm$ and $J_{zz}$ only has been studied
theoretically.\cite{PhysRevLett.100.047208,PhysRevB.69.064404} It does
indeed support a QSL ground state, for sufficiently small
$J_\pm/J_{zz}$.  For larger $J_\pm/J_{zz}$, the ground state is
instead a magnet with $\langle S_i^\pm \rangle \neq
0$.\cite{PhysRevLett.100.047208} While the actual value of
$J_\pm/J_{zz}\approx 0.3$ would place this model in the magnetic
state,\cite{PhysRevLett.100.047208} the $J_{z\pm}$ interaction in
particular is non-negligible in Yb$_2$Ti$_2$O$_7$, and preliminary
theoretical work suggests that it tends to stabilize the QSL state.
Indeed, in perturbation theory, the leading effect of the $J_{z\pm}$
coupling is to induce in the effective Hamiltonian a term close to the
Rokhsar-Kivelson interaction of Ref.~\onlinecite{PhysRevB.69.064404}
(see Supplementary Information), which was shown to stabilize the
QSL.\cite{PhysRevB.69.064404,rokhsar-kivelson} Although perturbation
theory is strictly speaking only valid for $J_{z\pm}/J_{zz}\ll1$, the
conclusions of this analysis are likely to extend to a larger
range of values.  A non-perturbative study of the full Hamiltonian in
Eq.~(\ref{eq:4}) is beyond the scope of this paper, but will be
reported in a future publication.  Given the uncertainty in the phase
boundary for the QSL state, we cannot disregard the possibility that
an ideal sample of \yto would be magnetically ordered, but that such order
is here disrupted by crystal defects.  This possibility should be pursued
further in the future taking into account the new understanding of the
Hamiltonian.  We proceed now to discuss the implications of the
alternative possibility of a zero field QSL state.

\bigskip

Many of the key properties of the QSL state of the $J_{zz}-J_\pm$ model
were established in Ref. \onlinecite{PhysRevB.69.064404}.  Conceptually, it is
a quantum analog of the classical regime of Coulombic correlations
observed in spin ice.\cite{fennell2009magnetic} Specifically, where spin
ice realizes an analog of magnetostatics, the QSL state of
Eq.~(\ref{eq:4}) embodies a complete fictitious {\sl quantum
  electrodynamics}.  In this phase, the magnetic monopoles of spin ice
become full-fledged coherent excitations of the system.  In addition,
the QSL supports dual {\sl electric monopoles} and a dynamical {\sl
  emergent photon} mode at low energy.  The complex and largely diffuse
character of the scattering in zero field\cite{ross2009} may well be a consequence of
the combination of these diverse excitations.  Indeed, where a neutron
can create just one spin wave ($S=1$) excitation, the $S=1/2$ magnetic
monopoles are excited in pairs with no {\sl individual} momentum
conservation constraints.  A careful comparison of 
theoretical modeling and focused experiments in low field is clearly needed.

A key consequence of the QSL scenario is the presence of a {\sl quantum
  confinement phase transition} in applied field (see Fig.~3). Such a
quantum phase transition is required to remove the ``fractional''
excitations of the QSL phase (electric and magnetic monopoles) from the
spectrum in the semiclassical high field phase.  Theoretically, such
quantum critical points have been studied in related model Hamiltonians,
and occur by a mechanism analogous to the Higgs transition in the
standard model\cite{Senthil2004a} or by magnetic monopole
condensation\cite{Bergman2006}.  The gapless excitations observed by
neutrons at $H\approx0.5$T in Ref.~\onlinecite{ross2009} indeed indicate a
quantum critical point at this field.  Further theoretical work is
required for detailed comparison to experiment.


\bigskip

\yto enjoys several major advantages over other materials considered
candidates for QSL states up to now:\cite{balents2010spin} it is the
only case in which the Hamiltonian parameters are precisely known, and
for which large single-crystal samples highly suitable for detailed
neutron scattering measurements are available.   Moreover, similar
methods may be applicable to other rare earth pyrochlores in which
substantial quantum effects are present, such as Er$_2$Ti$_2$O$_7$ and
Tb$_2$Ti$_2$O$_7$.\cite{gardner2010magnetic,molavian-gingras-canals}  Especially in this broader
context, the prospects for detailed observation of the long-sought QSL
physics are bright.  The basic framework established here will allow
coordinated theoretical and experimental studies to confront the
problem.

\oddsidemargin=-.0cm
\evensidemargin=-.0cm
\textwidth=16.5cm
\topmargin=-1.35cm
\textheight=22cm


\makeatletter
\renewcommand\@biblabel[1]{#1.}
\makeatother
\bibliography{Balents_bib}

\oddsidemargin=-.0cm
\evensidemargin=-.0cm
\textwidth=16.5cm
\topmargin=-1.35cm
\textheight=22cm


\noindent{\bf Acknowledgements}   We thank M. Gingras for abundant comments on a first draft.  K.A.R. acknowledges useful discussions with B. Prasanna Venkatesh.  L.S. acknowledges B. Canals and P. Holdsworth for many enlightening  discussions, and the Institut N\'{e}el, Grenoble for hospitality.  K.A.R. and B.D.G. were supported by NSERC of Canada.  L.B. and L.S. were supported by the DOE through Basic Energy Sciences  grant DE-FG02-08ER46524. 

\bigskip
 
\noindent{\bf Author Contributions} Experiments were performed by K.A.R. and B.D.G..  Theoretical calculations were done by L.S. and L.B..  The fitting was done jointly.  All authors were equally responsible for writing the manuscript.

\bigskip

\noindent{\bf Competing Interests}  The authors declare that they have no competing financial interests.
 

\clearpage

\oddsidemargin=-.0cm
\evensidemargin=-.0cm
\textwidth=16.5cm
\topmargin=-5cm
\textheight=24cm

\begin{figure}
\begin{center}    
\includegraphics[width=5.6in]{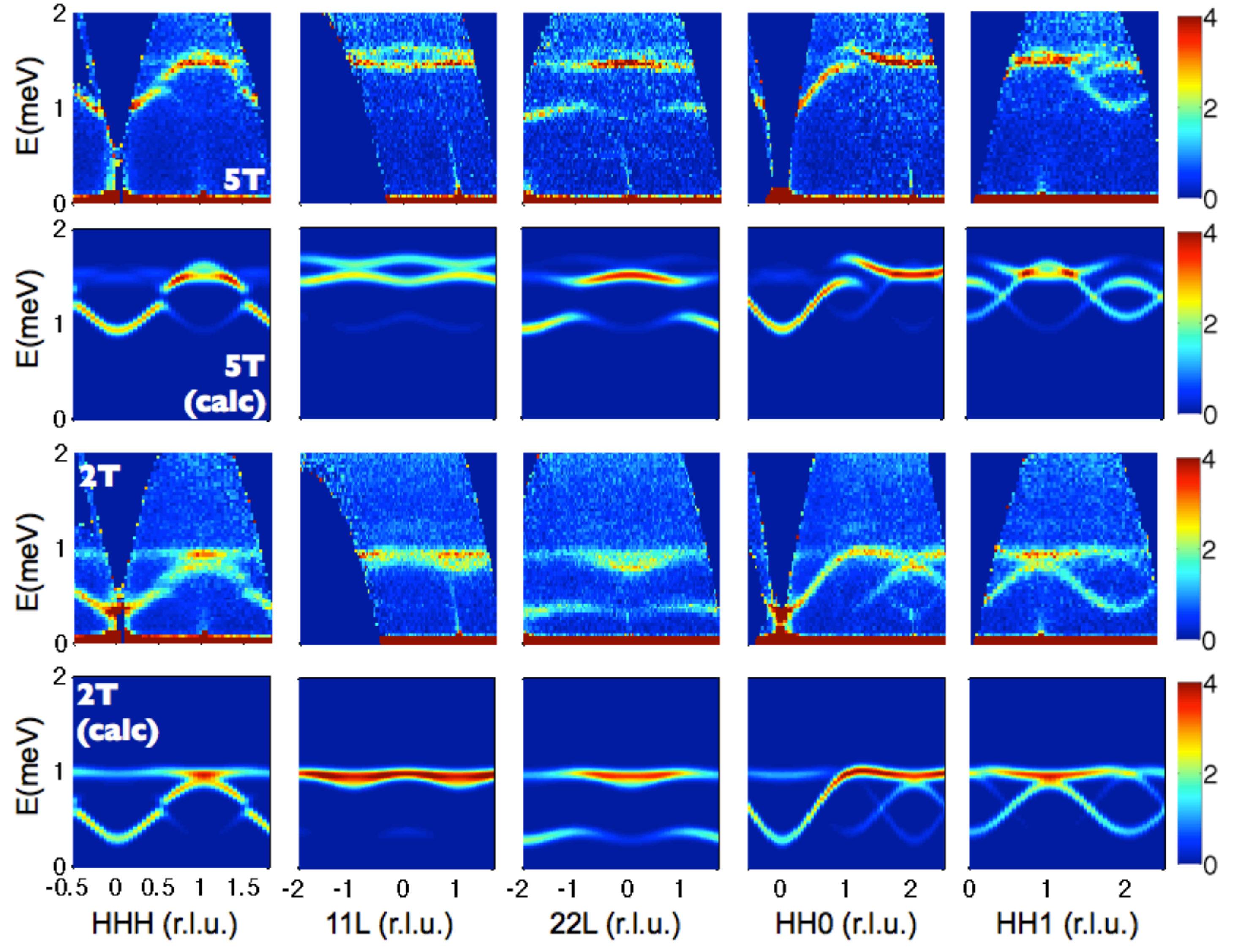}
\end{center}
\caption*{{\bf Figure 1:}   The measured $S({\bf Q},\omega)$ at $T=\;$30mK, sliced along various directions in the HHL plane, for both $H=\;$5T (first row) and $H=\;$2T (third row).   The second and fourth row show the calculated spectrum for these two field strengths, based on an anisotropic exchange model with five free parameters (see text) that were extracted by fitting to the 5T data set.  For a realistic comparison to the data, the calculated $S({\bf Q},\omega)$ is convoluted with a gaussian of full-width 0.09meV.  Both the 2T and 5T data sets, comprised of spin wave dispersions along five different directions, are described extremely well by the same parameters. }
     \label{fig:figure1}
\end{figure}
\begin{figure} 
\begin{center}
   \includegraphics[width=5.1in]{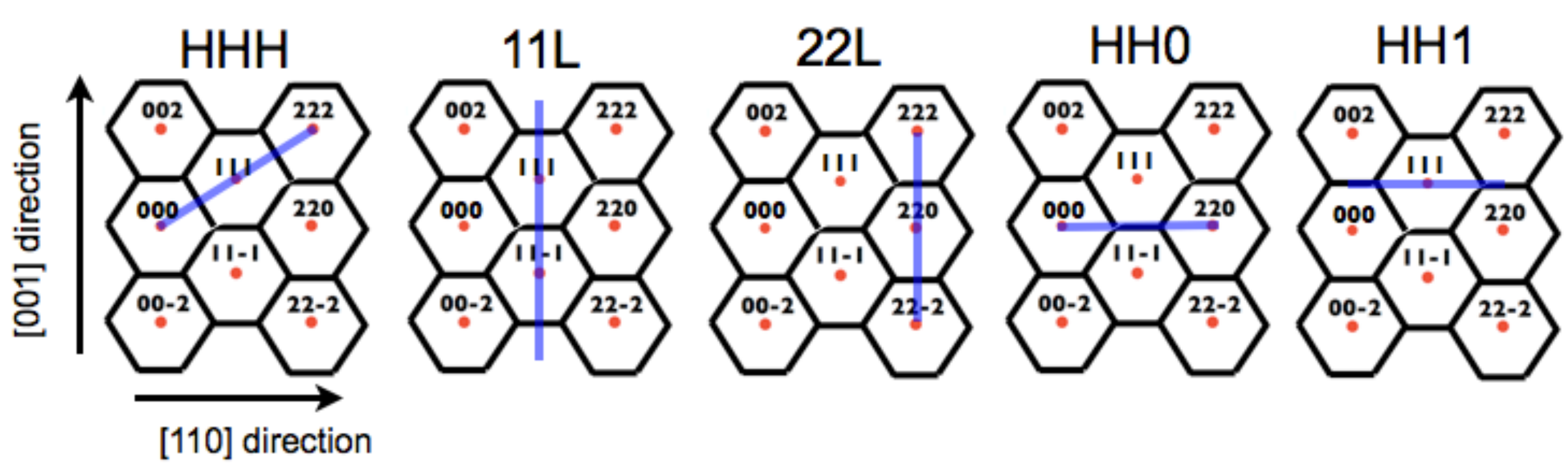} 
\end{center}
   \caption*{{\bf Figure 2:}   Representations of the HHL scattering plane, showing the FCC Brillouin zone boundaries and the corresponding zone centers (labelled in terms of the conventional simple-cubic unit cell).  Blue lines indicate the directions of the five different cuts shown in Figure 1.}
     \label{fig:figure2}
\end{figure}

\oddsidemargin=-.0cm
\evensidemargin=-.0cm
\textwidth=16.5cm
\topmargin=-1.35cm
\textheight=22cm

\clearpage

\begin{figure}
\begin{center}
   \includegraphics[width=4in]{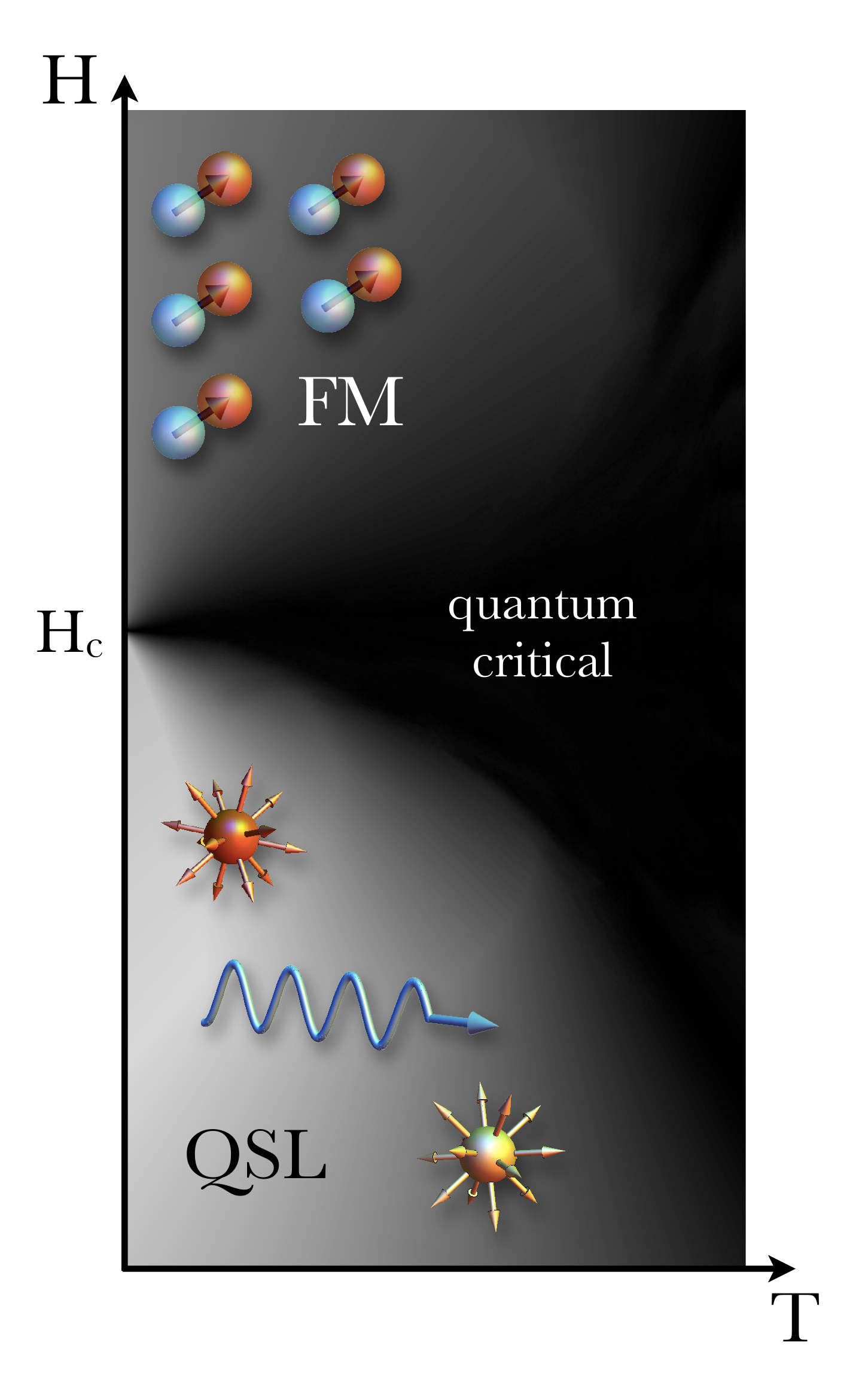} 
\end{center}
   \caption*{{\bf Figure 3:}   Schematic phase diagram in the temperature ($T$) -
       magnetic field ($H$) plane, for a material in the quantum spin
       liquid (QSL) phase of Eq.~(1) at $T=H=0$.  At low field
       and temperature, the QSL state supports exotic excitations:
       magnetic (red sphere) and ``electric'' (yellow sphere) monopoles,
       and an emergent photon (wavy line).  The field $H_c$ marks a
       quantum critical point: the confinement phase transition.  For
       $H>H_c$, the ground state is a simple field-polarized ferromagnet
       (FM), and the elementary excitations are conventional magnetic
       dipoles.  The gradations in gray scale indicate crossovers to the quantum
       critical region between them, governed by the $T=0$ confinement
       phase transition.}
\end{figure}

\clearpage

\maketitle

\section{Cubic Rotations and Local Bases}
\label{sec:symmetries}

As described in the main text, we use the usual coordinate system for the pyrochlore lattice, with sites located on tetrahedra whose centers form a FCC lattice. We take one to be centered at the origin with its four corners at $\mathbf{r}_0=\frac{a}{8}(1,1,1)$, $\mathbf{r}_1=\frac{a}{8}(1,-1,-1)$, $\mathbf{r}_2=\frac{a}{8}(-1,1,-1)$ and $\mathbf{r}_3=\frac{a}{8}(-1,-1,1)$. The exchange matrices $\mathbf{J}_{ij}$ between sites of types $i$ and $j$ are obtained by applying the following cubic rotations $\mathcal{R}_{ij}$ to $\mathbf{J}_{01}$:
\begin{itemize}
\item $\mathcal{R}_{02}$ is a $\frac{2\pi}{3}$ rotation about the $[111]$ axis,
\item $\mathcal{R}_{03}$ is a $\frac{4\pi}{3}$ rotation about the $[111]$ axis,
\item $\mathcal{R}_{21}$ is a $\frac{4\pi}{3}$ rotation about the $[1\bar{1}\bar{1}]$ axis,
\item $\mathcal{R}_{31}$ is a $\frac{2\pi}{3}$ rotation about the $[1\bar{1}\bar{1}]$ axis,
\item $\mathcal{R}_{23}$ is a rotation made of a $\frac{2\pi}{3}$ rotation about the $[111]$ axis followed by a $\frac{4\pi}{3}$ rotation about the $[1\bar{1}\bar{1}]$ axis.
\end{itemize}
Note $\mathbf{J}_{ji}=\mathbf{J}_{ij}^T$.

We use the following local $(\mathbf{\hat{a}}_i,\mathbf{\hat{b}}_i,\mathbf{\hat{e}}_i)$ bases 
\begin{equation}
\left\{\begin{array}{l}
\mathbf{\hat{e}}_0=\frac{1}{\sqrt{3}}(1,1,1)\\
\mathbf{\hat{e}}_1=\frac{1}{\sqrt{3}}(1,-1,-1)\\
\mathbf{\hat{e}}_2=\frac{1}{\sqrt{3}}(-1,1,-1)\\
\mathbf{\hat{e}}_3=\frac{1}{\sqrt{3}}(-1,-1,1),
\end{array}\right.,
\qquad
\left\{\begin{array}{l}
\mathbf{\hat{a}}_0=\frac{1}{\sqrt{6}}(-2,1,1)\\
\mathbf{\hat{a}}_1=\frac{1}{\sqrt{6}}(-2,-1,-1)\\
\mathbf{\hat{a}}_2=\frac{1}{\sqrt{6}}(2,1,-1)\\
\mathbf{\hat{a}}_3=\frac{1}{\sqrt{6}}(2,-1,1)
\end{array}\right.,
\end{equation}
$\mathbf{\hat{b}}_i=\mathbf{\hat{e}}_i\times\mathbf{\hat{a}}_i$, and the $4\times4$ complex unimodular matrices
\begin{equation}
\zeta=\left(\begin{array}{cccc}
0 & -1 & e^{i\frac{\pi}{3}} & e^{-i\frac{\pi}{3}}\\
-1 & 0 & e^{-i\frac{\pi}{3}} & e^{i\frac{\pi}{3}}\\
e^{i\frac{\pi}{3}} & e^{-i\frac{\pi}{3}} & 0 & -1\\
e^{-i\frac{\pi}{3}} & e^{i\frac{\pi}{3}} & -1 & 0
\end{array}\right),
\quad \gamma=-\zeta^*,
\end{equation}
for which our exchange Hamiltonian takes the form of Eq.~(4).


\section{Curie-Weiss Temperature}


A high-temperature expansion of the Hamiltonian of Eq.~(1)
yields the $O(1/T^2)$ term in the uniform susceptibility from which we
extract the Curie-Weiss temperature,
\begin{eqnarray}
  \label{eq:CW-supplmat}
&&  \Theta_{CW} =  \frac{-1}{6k_B(2g_{xy}^2+g_z^2)} \big[ 2g_{xy}^2(4J_1
  + J_2 - 5J_3+2J_4) \\
  && + 8 g_{xy}g_z (J_1+J_2+J_3-J_4) + g_z^2(2J_1-J_2+2J_3+4J_4)\big],\nonumber
\end{eqnarray}
where $k_B$ is the Boltzmann constant.  Using the formulation of Eq.~(4) $\Theta_{CW}$ takes the simpler form 
\begin{equation}
  \label{eq:CW-supplmat-simple}
\Theta_{CW} =  \frac{1}{2k_B(2g_{xy}^2+g_z^2)} \big[\,g_z^2J_{zz} -4g_{xy}^2(J_\pm+2J_{\pm\pm})-8\sqrt{2}\,g_{xy}\,g_z J_{z\pm}\,\big].
\end{equation}

\section{Spin Wave Theory}
\label{sec:spinwaves}

As usual, we expand the Hamiltonian about one of the
classical states using Holstein-Primakoff bosons in the spirit of large
$s$, and keep only terms of and up to order $s$, which shall then be set
to $1/2$. We define the transverse Holstein-Primakoff bosons
$x_a=x_a^\dagger$, $y_a=y_a^\dagger$, conjugate with one another on site
$a$ of the pyrochlore lattice, satisfying
\begin{equation}
[x_a,y_a]=i, \qquad n_a=\frac{x_a^2+y_a^2}{2}-\frac{1}{2},
\end{equation}
such that
\begin{equation}
\mathbf{S}_a\cdot\mathbf{u}_a=s-n_a,\quad\mathbf{S}_a\cdot\mathbf{v}_a=\sqrt{s}\,x_a,\quad\mathbf{S}_a\cdot\mathbf{w}_a=\sqrt{s}\,y_a \;.
\end{equation}
Here $(\mathbf{v}_a,\mathbf{w}_a,\mathbf{u}_a)$ is an orthonormal basis,
chosen so that $\mathbf{u}_a$ gives the direction of the spin in the
classical ground state at site $a$ (which we find numerically).  We find
that for all fields, the ground state does not enlarge the unit cell, so
that there are only four distinct such bases, which we denote by
$a=0,..,3$.  One may choose, for example,
$\mathbf{v}_a=\mathbf{u}_a\times(1,1,1)/\|\mathbf{u}_a\times(1,1,1)\|$
and $\mathbf{w}_a=\mathbf{u}_a\times\mathbf{v}_a$.

Since the classical ground state does not enlarge the unit cell, we can
readily proceed to Fourier space in the four site basis.  Keeping only
terms of order $s$, we arrive at the spin-wave (quadratic) Hamiltonian,
\begin{equation}
H_{\mathbf{k}}=
\left(\begin{array}{cc} X_{-\mathbf{k}}^T & Y_{-\mathbf{k}}^T \end{array}\right)
\left(\begin{array}{cc}
A_{\mathbf{k}} & C_{\mathbf{k}} \\
C_{\mathbf{k}}^T & B_{\mathbf{k}}
\end{array}\right)
\left(\begin{array}{c} X_{\mathbf{k}} \\ Y_{\mathbf{k}} \end{array}\right),
\end{equation}
where $\left(\begin{array}{cc} X^T & Y^T \end{array}\right)=\left(\begin{array}{cccccc} x_0 & .. & x_3 & y_0 & .. & y_3 \end{array}\right)$ and $D_\mathbf{k}^{ab}=\tilde{D}_{ab}\cos(\mathbf{k}\cdot(\mathbf{r}_a-\mathbf{r}_b))$, $D=A,B,C$, with
\begin{eqnarray}
\tilde{A}_{ab}&=&s\mathbf{v}_a\cdot\mathbf{J}_{ab}\cdot \mathbf{v}_b +
\frac{\mu_B}{2}\mathbf{H}\cdot \mathbf{g}_a\cdot\mathbf{u}_a \,\delta_{ab}
-s\mathbf{u}_a\cdot\sum_b \mathbf{J}_{ab}\cdot\mathbf{u}_b \nonumber\\
\tilde{B}_{ab}&=&s\mathbf{w}_a\cdot\mathbf{J}_{ab}\cdot \mathbf{w}_b +
\frac{\mu_B}{2}\mathbf{H}\cdot \mathbf{g}_a\cdot\mathbf{u}_a\,\delta_{ab} -s\mathbf{u}_a\cdot\sum_b \mathbf{J}_{ab}\cdot\mathbf{u}_b \nonumber\\
\tilde{C}_{ab}&=&s \mathbf{v}_a\cdot \mathbf{J}_{ab}\cdot \mathbf{w}_b,
\end{eqnarray}
where $\mathbf{H}$ is the magnetic field and $\mathbf{J}_{ab}$ and $\mathbf{g}_a$ are $3\times3$ matrices with matrix elements $J_{ab}^{\mu\nu}$ and $g_{a}^{\mu\nu}$, respectively.  To find the modes, we resort to the path-integral formulation. The action at temperature $T=1/(k_B\beta)$ is 
\begin{equation}
\mathcal{S}=\frac{1}{2\beta}\sum_n\sum_{\mathbf{k}}
Z^T_{-\mathbf{k},-\omega_n}
\left[G_{\mathbf{k}}+(i\omega_n)\Gamma\right]
Z_{\mathbf{k},\omega_n},
\end{equation}
where $\omega_n=\frac{2\pi n}{\beta}$ is the Matsubara frequency, where we have defined $Z^T=\left(\begin{array}{cc} X^T & Y^T \end{array}\right)$,
\begin{equation}
G_{\mathbf{k}}=
2\left(\begin{array}{cc}
A_{\mathbf{k}} & C_{\mathbf{k}} \\
C_{\mathbf{k}}^T & B_{\mathbf{k}}
\end{array}\right)\quad\mbox{and}\quad
\Gamma=
\left(\begin{array}{cc}
0 & -i\mathbf{1}_4 \\
i\mathbf{1}_4 & 0
\end{array}\right)
\end{equation}
($\mathbf{1}_4$ is the four-by-four identity matrix). As usual, the {\sl real frequency} dispersion relations $\omega(\mathbf{k})$ are found by solving the zero eigenvalue equations of the matrix $G_{\mathbf{k}}+\omega\Gamma$.  Here, these are equivalently the (both zero and non-zero) eigenvalues of $-\Gamma G_\mathbf{k}$.

We calculate the inelastic structure factor (to which the intensity $I(\mathbf{k},\omega)$ of the scattering is proportional) obtained from the moment-moment correlation function,\cite{neutronscattbook}
\begin{equation}
S(\mathbf{k},\omega)=\sum_{\mu,\nu}\left[\delta_{\mu\nu}-(\mathbf{\hat{k}})_\mu(\mathbf{\hat{k}})_\nu\right]\sum_{a,b}\langle m_a^\mu(-\mathbf{k},-\omega)m_b^\nu(\mathbf{k},\omega)\rangle, 
\end{equation}
where $\mathbf{\hat{k}}$ is the unit vector associated with
$\mathbf{k}$, $m_a^\mu(\mathbf{k},\omega)=\mu_B
\sum_{\kappa}g_a^{\mu\kappa}S_a^\kappa(\mathbf{k},\omega)$ is the moment
at momentum $\mathbf{k}$ and real frequency $\omega$ on the $a=0,..,3$
sublattice in direction $\mu=x,y,z$, and $S_a^\kappa$ is, as usual, the
$\kappa$th coordinate of the effective spin-1/2 spin at site $a$. $\mathcal{F}_{\mu\nu}(\mathbf{k})=\delta_{\mu\nu}-(\mathbf{\hat{k}})_\mu(\mathbf{\hat{k}})_\nu$ selects the component of the spin-spin correlations perpendicular to the scattering vector,\cite{neutronscattbook} and $\mathcal{G}^{\mu\nu}_{ab}(\mathbf{k},\omega)=\langle m_a^\mu(-\mathbf{k},-\omega)m_b^\nu(\mathbf{k},\omega)\rangle$ is the moment-moment correlation function that originates from the interaction between the neutrons' moment and the spins' moments, which we find to be  
\begin{equation}
\mathcal{G}^{\mu\nu}_{ab}=-\pi s\eta_{a}^\mu\eta_b^\nu\sum_\alpha \delta(\omega-\epsilon_\alpha)
  \frac{1}{
    \psi_{R,\alpha}^\dagger \Gamma \psi_{R,\alpha}}
  \left([{\psi}_{R,\alpha}^{\dagger}]_a [{\psi}_{R,\alpha}^{\vphantom\dagger}]_b\right),
\end{equation}
where $\delta$ is the Dirac distribution, $\eta_a^\mu=\sum_{\kappa=x,y,z} g_a^{\mu\kappa}\left(\begin{array}{cc} v_a^\kappa & w_a^\kappa \end{array}\right)$, $\epsilon_\alpha$ is the $\alpha$th eigenvalue of $-\Gamma G$, and  $\psi_{R,\alpha}$ is its corresponding ``right'' eigenvector, i.e. such that $-\Gamma G\cdot \psi_{R,\alpha}=\epsilon_\alpha\,\psi_{R,\alpha}$. Also note that the momentum and frequency dependences are implied everywhere to be $\mathbf{k},\omega$.

Now we estimate the amplitude of quantum fluctuations using our spin
wave theory, by evaluating the quantum moment reduction in zero field
and at zero temperature.  In the Holstein-Primakoff boson language the
reduction of the spin expectation value from the classical value of
$1/2$, averaged over the four site basis, is
$r=\frac{1}{4}\sum_{a=0}^3 \langle 0| n_a|0\rangle$, where
$\langle0|n_a|0\rangle$ indicates a ground state quantum expectation
value.  Evaluating this using the path integral method, we obtain
$r=\frac{1}{8}\int_{\mathbf{k},\omega}\mbox{Tr }\langle
\left[G_{\mathbf{k}}+\omega\Gamma\right]^{-1}\rangle-\frac{1}{2}$, i.e.
\begin{equation}
r=\frac{1}{8}\int_{\mathbf{k}}\sum_\alpha\Theta(\epsilon_\alpha)\frac{\psi_{R,\alpha}^\dagger\psi_{R,\alpha}}{\psi_{R,\alpha}^\dagger\Gamma\psi_{R,\alpha}}-\frac{1}{2}\approx0.05\,
\end{equation}
($\Theta$ is the usual Heaviside distribution), which is a 10\% reduction compared with the classical value of $1/2$.

\section{Further Details of Experimental Data and Fits}
\label{sec:exp-fit}

The inelastic neutron scattering data (rows 1 and 3 in Figure 1) contains several features worth commenting on further.  First, the darkest blue areas do not contain any data, either for kinematic reasons near $\mathbf{Q}=(0,0,0)$, or because of the finite angular extend of the scan.  Second, at $E=0$ one observes intensity which is due to coherent and incoherent elastic scattering from the sample, and hence is more intense than the inelastic features by up to several orders of magnitude, thus appearing red (off scale).  Third, near the $(0,0,0)$ position there is higher background leading to unphysical intensities due to contamination from the un-scattered incident beam.  This is observable in the HHH and HH0 data sets near the zero position.  

The data sets were collected by counting for 8 minutes per angular rotation of the sample.  The total time per magnetic field setting was about 12 hours.

Supplementary Figure 1 shows the dispersions obtained from the fitting procedure over-plotted on the data.  These fits were accomplished by digitizing the shape of the dispersion from the experimental data, and performing a least squares minimization routine to match it.   Intensities were calculated based on the spin wave theory using the extracted exchange parameters, and were not fit to the data.  The 5D parameter space (four exchange plus one g-tensor parameter) was explored using a uniform search technique with the resulting excellent description obtained (Figure 1).
\begin{figure}[htbp]
\begin{center}
\includegraphics[scale=.46]{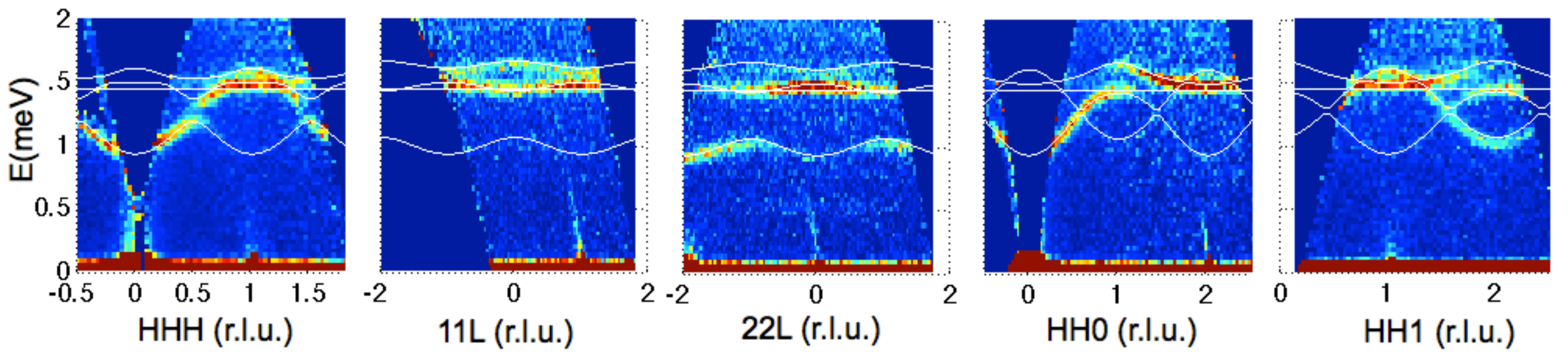} %
\caption*{{\bf Supplementary Figure 1:}  Dispersions (white curves) obtained from the fitting procedure over-plotted on the data at $H=5\,$T.}
\label{fig:values}
\end{center}
\end{figure}

\section{Features of the Spin Wave Spectrum}
\label{sec:features}

The spectra of Figure 1 are very rich.  Because of the many parameters
and of the presence of {\sl classical} phase transitions (there are
several ground states to the classical model), it is very difficult to
track which features are due to what terms of Eq.~(4).  One remarkable
feature which can be easily identified throughout most of phase space
(including the region around which Eq.~(3) places Yb$_2$Ti$_2$O$_7$) is a
quasi-flat band.  Specifically, one spin wave mode is completely
dispersionless in the plane of reciprocal space with $k_x=k_y$,
i.e. normal to the magnetic field direction and passing through the
origin in reciprocal space.  All scattering measurements on \yto have
been taken in this plane, so this feature is quite significant in the
experiments.    In the region of phase space around \yto and for $H=5\,$T,
we find its energy to be, numerically:
\begin{equation}
E_{{\rm 2d\, flat}}\,\approx\, 0.74+0.51\, J_{zz}-1.18\, J_{\pm}-3.11\, J_{\pm\pm}-5.81\, J_{z\pm}\quad\mbox{meV},
\end{equation}
where the $J_\alpha$'s must be input in meV, and which for our fit gives $E_{{\rm 2d\,
    flat}}^{{\mbox{\tiny{Yb$_2$Ti$_2$O$_7$}}}}\approx 1.45\,$meV.
Note that the energy of this feature is most sensitive to $J_{z\pm}$.

Moreover, we observe the following trends in the region around Yb$_2$Ti$_2$O$_7$:
\begin{itemize}
\item as $|J_{z\pm}|$ increases, all bands go up in energy (especially the two-dimensional flat one),
\item increases in $J_\pm$ and $J_{\pm\pm}$ seem to have more or less the same effect: the bands get closer, and this happens in particular because the energy of the top bands decreases.
\end{itemize}

\section{Mean Field Theory Calculation}
\label{sec:MFT}


The Curie-Weiss mean field Hamiltonian obtained from Eq.~(1) takes
the form
\begin{eqnarray}
H_{\mbox{{\tiny MF}}}&=&\sum_{\langle i,j\rangle}\sum_{\mu,\nu}J_{ij}^{\mu\nu}\left(\langle S_i^\mu\rangle S_j^\nu+S_i^\mu\langle S_j^\nu\rangle - \langle S_i^\mu\rangle\langle S_j^\nu\rangle\right)\nonumber\\
&&- \mu_B H^\mu \sum_i g_i^{\mu \nu} S_i^\nu,
\label{eq:MFHam}
\end{eqnarray}
where $H^\mu$ is the magnetic field in the $\mu$ direction, $\langle S_i^\mu\rangle$ is the mean field quantum thermal expectation value of $S_i^\mu$, defined by $\langle S_i^\mu\rangle=\frac{1}{Z}\mbox{Tr}\,S_i^\mu \exp(-\beta H_{\mbox{{\tiny MF}}})$, where $\beta$ is the inverse temperature $\beta=1/(k_B T)$, where $k_B$ is the Boltzmann constant and $Z$ is the partition function $Z=\mbox{Tr}\,\exp(-\beta H_{\mbox{{\tiny MF}}})$. The traces are taken over the up and down spin states of every spin. 

The ground state does not enlarge the unit cell at zero temperature, and we assume this is still the case at non-zero temperature. Thus, we define, for every sublattice $a$ and every axis $\mu$ the average magnetization 
\begin{equation}
m_a^\mu=\langle S_{t,a}^\mu\rangle,
\end{equation}
which is the same for every tetrahedron $t$, which can be either ``up''
or ``down''.  We arrive at the twelve consistency equations
\begin{equation}
\mathbf{m}_a=-\frac{\mathbf{h}_a^{\mbox{{\tiny eff}}}}{2\|\mathbf{h}_a^{\mbox{{\tiny eff}}}\|}\tanh\frac{\beta\|\mathbf{h}_a^{\mbox{{\tiny eff}}}\|}{2},\label{eq:6}
\end{equation}
where $\mathbf{h}_a^{\mbox{{\tiny eff}}}=2\sum_b\mathbf{m}_b\cdot \mathbf{J}_{ba}-\mu_B\mathbf{H}\cdot\mathbf{g}_a$; the free energy per site is
\begin{equation}
f=-\frac{1}{4}\sum_{a,b}\mathbf{m}_a\cdot\mathbf{J}_{ab}\cdot\mathbf{m}_b-\frac{1}{4\beta}\sum_a\ln\left[2\cosh\frac{\beta\|\mathbf{h}_a^{\mbox{{\tiny eff}}}\|}{2}\right].\label{eq:7}
\end{equation}


We solve Eqs.~(\ref{eq:6}) numerically, choosing the solution which
minimizes the free energy in Eq.~(\ref{eq:7}). With the field oriented
along the [$110$] direction, we find the phase diagram shown in Supplementary Figure~2, which contains two phases.  In zero field, the low temperature phase is a ferromagnet with net magnetization along any of
the $\langle$100$\rangle$ axes.  In low fields, this expands into a
phase in which the average magnetization is {\sl not} aligned with
the field, but lies within a \{001\} plane.  On increasing fields, a
transition occurs to the high field state in which the net magnetization
is aligned with the applied field, and no symmetries are spontaneously
broken.  This state is continuously connected to the high temperature
paramagnetic phase.


\begin{figure}[htbp]
\begin{center}
\includegraphics[scale=.75]{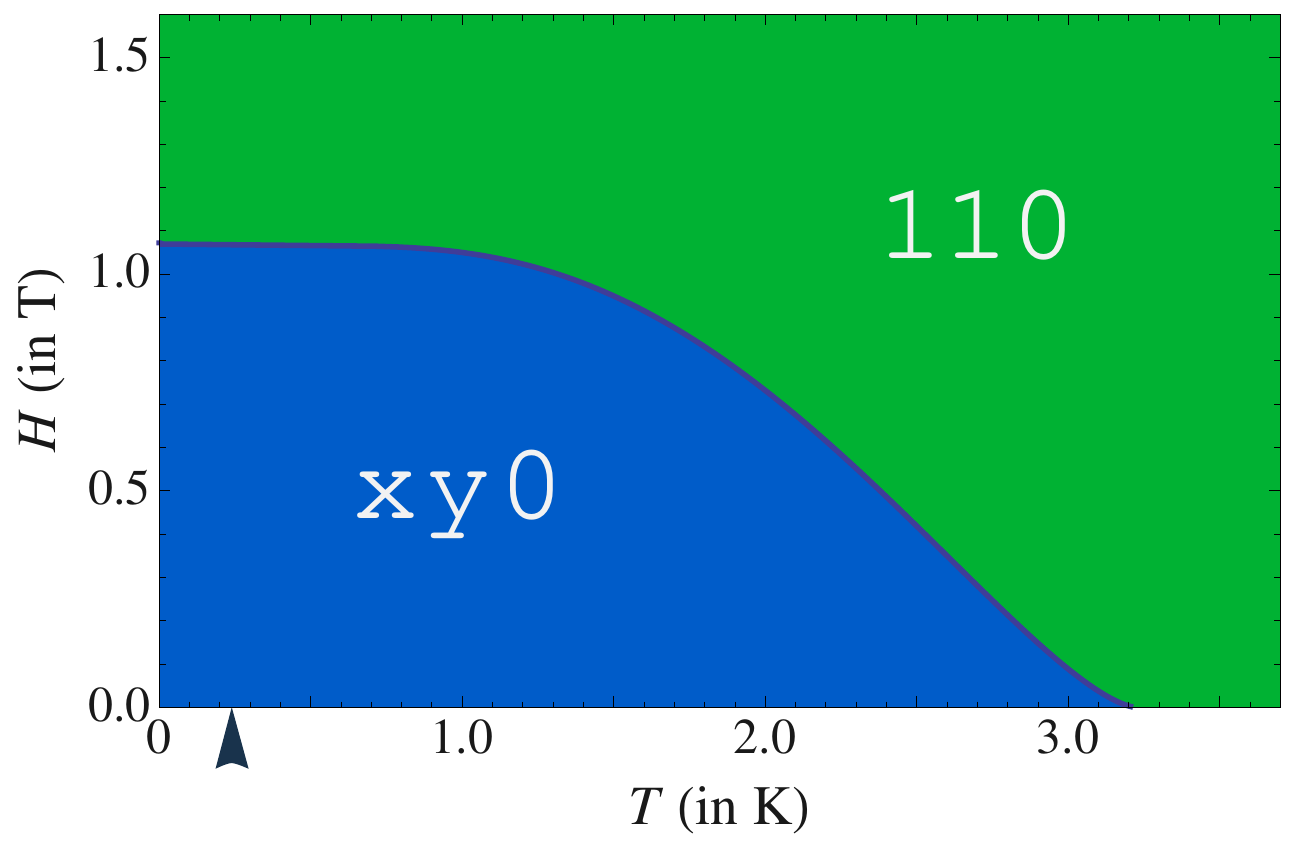} %
\caption*{{\bf Supplementary Figure 2:} Field versus temperature phase diagram obtained from a mean-field analysis of the Hamiltonian of Eq.~(1), for a field $\mathbf{H}$ parallel to the $[110]$ direction and with the exchange constant values obtained with our fits, Eq.~(3).  The system displays a net magnetic moment throughout the $(H,T)$ plane.  The blue region denotes a region where the total magnetization lies in the $xy$ plane, and the green region is the paramagnetic phase; the two zones are separated by a continuous transition.  In zero field, this transition takes place at $T_c^{MF}=3.2\,$K. At zero temperature, the amplitude of the transition field is $H_c^{MF}=1.1\,$T.  The dark blue arrow shows the experimentally reported transition temperature of $240\,$mK: the actual transition occurs at a much lower temperature than that predicted by mean-field theory $T_c^{MF}$.}
\label{fig:MFTdiag}
\end{center}
\end{figure}

\section{Perturbation Theory}
\label{sec:PT}


We show that the $U(1)$ QSL described by Hermele {\sl et al.}\cite{PhysRevB.69.064404,PhysRevLett.100.047208} is stable to the addition of the terms in the symmetry-obtained Hamiltonian Eq.~(4), provided all coupling constants are small with respect to the Ising exchange parameter $J_{zz}$.  In other words, we show that the $U(1)$ QSL exists in a finite region of parameter space, specifically where $J_\pm$, $J_{z\pm}$ and $J_{\pm\pm}$ are small with respect to $J_{zz}$.  

In that limit, one may apply perturbation theory.  When $J_{zz}>0$, the ground state manifold of the unperturbed Hamiltonian is the extensively degenerate ``two-in-two-out'' manifold.  When $J_{z\pm}=J_{\pm\pm}=0$,  Eq.~(4) can be mapped exactly onto the Hamiltonian of Ref.~\onlinecite{PhysRevB.69.064404}, where the first non-vanishing and non-constant term in perturbation theory (above the ``two-in-two-out'' manifold) was shown to be third order in $J_\pm/J_{zz}$:
\begin{equation}
H_{{\rm ring}}^{{\rm eff}}=-K\sum_{\{i,j,k,l,m,n\}=\hexagon}\left(\mathsf{S}_i^+\mathsf{S}_j^-\mathsf{S}_k^+\mathsf{S}_l^-\mathsf{S}_m^+\mathsf{S}_n^-+\mbox{h.c.}\right),
\end{equation}
where $K=\frac{12J_\pm^3}{J_{zz}^2}$ is a ring exchange interaction.
$H_{{\rm ring}}^{{\rm eff}}$ flips the spins on the ``flippable''
hexagons, i.e. those with alternating up and down spins, and yields zero
otherwise. This represents a ring move responsible for favoring the
quantum superpositions of the $U(1)$ QSL.    As shown in
Refs.~\onlinecite{PhysRevB.69.064404,PhysRevLett.100.047208,shannon2011quantum},
this ring Hamiltonian has as its ground state a $U(1)$ QSL, whose low
energy physics is described as the Coulomb phase of a $U(1)$ gauge
theory.  This phase is locally stable to {\sl all} perturbations\cite{PhysRevB.69.064404} in
three dimensions, which is enough already to guarantee the persistence
of the QSL state when the other exchange couplings are sufficiently
small, i.e. when the induced terms in the effective Hamiltonian are much
smaller than the ring coupling $K$.

We can, however, go further and consider these effects explicitly in the
perturbative limit, which extends the discussion to the case when
$J_{z\pm}, J_{\pm\pm} \ll J_{zz}$ but with no particular assumptions
placed upon the magnitude of the induced terms in the effective
Hamiltonian relative to $K$.  When $J_{z\pm}$ or $J_{\pm\pm}$ are
non-zero, other ``non-ring'' effective Hamiltonians are allowed.  In
particular, 
the $J_{z\pm}$ term gives rise to an effective third neighbour
ferromagnetic Ising Hamiltonian:
\begin{equation}
H_{{\rm 3rd\, Ising}}^{{\rm eff}}=-J_{(3)}\sum_{\langle\langle\langle i,j\rangle\rangle\rangle}\mathsf{S}_i^z\mathsf{S}_j^z,
\end{equation}
where $J_{(3)}=\frac{3J_{z\pm}^2}{J_{zz}}$.  This term alone has six symmetry-related ordered ground states.  Each of
them consists of the choice of one of the six two-in-two-out tetrahedra,
with the {\sl same} pattern repeated on each ``up'' tetrahedron (these
states are magnetic with their moment along the $\langle100\rangle$
directions) and contain no flippable hexagons, see Supplementary Figure
3.  
\begin{figure}[htbp]
\begin{center}
\includegraphics[scale=.42]{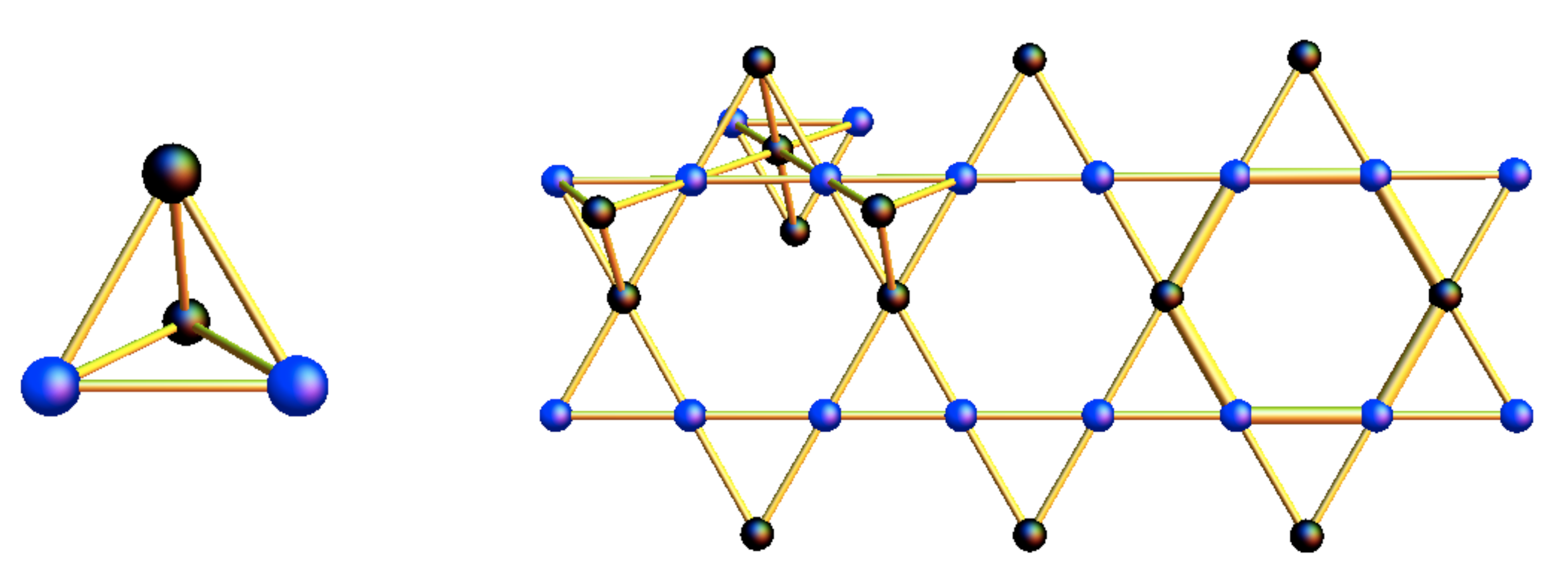} %
\caption*{{\bf Supplementary Figure 3:} One of the six ground states of $H_{{\rm 3rd\, Ising}}^{{\rm eff}}=-\frac{3J_{z\pm}^2}{J_{zz}}\sum_{\langle\langle\langle i,j\rangle\rangle\rangle}\mathsf{S}_i^z\mathsf{S}_j^z$.  A blue sphere represents an ``in'' spin while a black sphere represents an ``out'' one.  Because two of the four chain types contain {\sl non}-alternating spins, there are no flippable hexagons in any of the six (equivalent) ground states.}
\label{fig:FMstate}
\end{center}
\end{figure}
This implies that states which contain flippable hexagons represent
an energy cost.  We can therefore consider, $H_{{\rm 3rd\, Ising}}^{{\rm eff}}$
as an analog of the Rokhsar-Kivelson term introduced by Hermele {\sl et
  al.},\cite{rokhsar-kivelson,PhysRevB.69.064404} $H_{RK}=VN_{{\rm f}}$, where $N_{{\rm f}}$ is the
operator that counts the number of flippable hexagons.  This term
actually {\sl stabilizes} the QSL
state\cite{PhysRevB.69.064404,shannon2011quantum}.  In particular, the
QSL phase grows in stability as $V$ is increased from zero up to the
point $V=K$, beyond which (for $V>K$), the system undergoes a first
order transition to a degenerate set of classical unflippable states,
which includes the six ordered $\langle 100\rangle$ ferromagnetic ground
states described above.  

In the case relevant here, when $J_{(3)}/K$ is sufficiently large, the
system must undergo a transition to the unflippable $\langle
100\rangle$ ground states.  Since this model and the RK one of
Ref.~\onlinecite{PhysRevB.69.064404} agree on the phases both when
$J_{(3)} \ll K$ and when $J_{(3)} \gg K$, it is natural to expect that
the intervening phase diagram coincides in the two models as well.
Therefore we expect that the QSL state is maximally stable when
$J_{(3)}/K$ takes some value of $O(1)$.  (For the values of the
exchange constants given by our fits we find
$J_{(3)}/K=\frac{J_{z\pm}^2J_{zz}}{4J_{\pm}^3}=6.2$, but we caution
that this is probably outside the perturbative regime).  We note in passing that the
$J_{(3)}$ exchange can also be expressed in terms of purely plaquette
interactions, which might allow further analytical connection to the
RK theory.  We will not, however, pursue this further here.

Inclusion of the other coupling $J_{\pm\pm}$, higher order effects,
and cross terms amongst the exchange couplings does not lead to any new
effects.  Indeed, all the associated terms in the effective Hamiltonian assume a
ferromagnetic or ring form, and can be subsumed in the above couplings.
They are also higher order in $J_\alpha/J_{zz}$,
$\alpha=\pm,z\pm,\pm\pm$.

\section{Comparison between our Exchange Constants and those of Thompson {\sl et al.}\cite{thompson2010rods}}
\label{sec:thompson-comp}


The correspondence between our effective spin-1/2 operators
$\mathbf{S}_i$ in Eq.~(1) and the full 7/2-angular-momentum
operators $\mathbf{J}_i$ used by Thompson {\sl et al.} in
Ref. \onlinecite{thompson2010rods} is given by projecting the full angular
momentum into the ground state Kramer's doublet
\begin{equation}
P_{1/2}\,\mathbf{J}_i\, P_{1/2}= \frac{\mathbf{g}_i\cdot\mathbf{S}_{i}}{g_J},
\end{equation}
where $P_{1/2}$ is the projection operator to the ground state Kramer's
doublet, and $g_J=8/7$ is the Land\'{e} factor.

We use $g_{xy}/g_z=2.4$ and $g_z=1.79$ for concreteness, but the
results do not depend too much upon the details of this choice within
the range of parameters found in the literature.  With this choice, we
find that the semi-formal relations between our parameters and those
given in Ref. \onlinecite{thompson2010rods} are
\begin{eqnarray}
J_1&=&0.818\,\mathcal{J}_{\mbox{{\tiny Ising}}}-9.08\,\mathcal{J}_{\mbox{{\tiny iso}}}-1.21\,\mathcal{J}_{\mbox{{\tiny pd}}}-2.34\,\mathcal{J}_{\mbox{{\tiny DM}}}\nonumber\\
J_2&=&-0.818\,\mathcal{J}_{\mbox{{\tiny Ising}}}-8.03\,\mathcal{J}_{\mbox{{\tiny iso}}}-11.2\,\mathcal{J}_{\mbox{{\tiny pd}}}+6.16\,\mathcal{J}_{\mbox{{\tiny DM}}}\nonumber\\
J_3&=&0.818\,\mathcal{J}_{\mbox{{\tiny Ising}}}+4.88\,\mathcal{J}_{\mbox{{\tiny iso}}}+12.7\,\mathcal{J}_{\mbox{{\tiny pd}}}-2.34\,\mathcal{J}_{\mbox{{\tiny DM}}}\nonumber\\
J_4&=&0.818\,\mathcal{J}_{\mbox{{\tiny Ising}}}-0.523\,\mathcal{J}_{\mbox{{\tiny iso}}}-5.49\,\mathcal{J}_{\mbox{{\tiny pd}}}+5.62\,\mathcal{J}_{\mbox{{\tiny DM}}}.\nonumber
\end{eqnarray}
For $\mathcal{J}_{\mbox{{\tiny Ising}}}=6.98\;10^{-2}\,$meV,
$\mathcal{J}_{\mbox{{\tiny iso}}}=1.90\;10^{-2}\,$meV,
$\mathcal{J}_{\mbox{{\tiny pd}}}=-2.50\;10^{-2}\,$meV and
$\mathcal{J}_{\mbox{{\tiny DM}}}=-2.33\;10^{-2}\,$meV as calculated by
Thompson {\sl et al.} in Ref.\onlinecite{thompson2010rods}, we obtain
$J_1=-0.03\,$meV, $J_2=-0.07\,$meV, $J_3=-0.11\,$meV and
$J_4=0.05\,$meV, or, in the formulation of Eq.~(4), $J_{zz}=1.9\;10^{-4}\,$meV, $J_{\pm}=3.0\;10^{-3}\,$meV, $J_{\pm\pm}=3.8\;10^{-2}\,$meV and $J_{z\pm}=-6.4\;10^{-2}\,$meV.  These values are rather different from those found in
our fits, Eq.~(3) and Eq.~(5), and indeed yield spin wave spectra
in strong disagreement with experiment.  In principle, one resolution of
the difference could be that the exchange couplings are actually
strongly temperature dependent, and distinctly different in the
temperature range studied in Ref.~\onlinecite{thompson2010rods} and at the low
temperatures studied here.  Such a change in exchange parameters could
conceivable occur if the transition observed at 214-240mK had a
substantial structural component.  However, we do not have any a priori
reason to suspect this.  In any case, it would be very interesting to
resolve the differences between the two exchange models.


\end{document}